\DocumentMetadata{lang=en, pdfversion=2.0, pdfstandard=ua-2, testphase={phase-III,firstaid,math,title}}
\documentclass[sigconf]{acmart-tagged}

\AtBeginDocument{%
  }

\copyrightyear{2026}
\acmYear{2026}
\setcopyright{cc}
\setcctype{by}
\acmConference[ASSETS '26]{The 28th International ACM SIGACCESS Conference on Computers and Accessibility}{October 25--28, 2026}{Vila Nova de Gaia, Portugal}
\acmBooktitle{The 28th International ACM SIGACCESS Conference on Computers and Accessibility (ASSETS '26), October 25--28, 2026, Vila Nova de Gaia, Portugal}
\acmDOI{10.1145/3797867.3841235}
\acmISBN{979-8-4007-2521-0/2026/10}

\begin{document}

\title{From Anonymous Shapes to Named Places: A Tool for Braille and Place-Semantic Annotation of Tactile Maps}

\author{Li Liu}
\orcid{0000-0002-5184-054X}
\email{lliu112@ucsc.edu}
\affiliation{%
  \institution{University of California, Santa Cruz}
  \city{Santa Cruz}
  \state{California}
  \country{USA}
}

\author{Ashmita Dua}
\orcid{0009-0007-8337-7848}
\email{asdua@ucsc.edu}
\affiliation{%
  \institution{University of California, Santa Cruz}
  \city{Santa Cruz}
  \state{California}
  \country{USA}
}

\author{Jiaming Qu}
\orcid{0000-0003-4460-5637}
\authornote{Work unrelated to position at Amazon.}
\email{qjiaming@amazon.com}
\affiliation{%
  \institution{Amazon}
  \city{Seattle}
  \state{Washington}
  \country{USA}
}

\author{David T. Lee}
\orcid{0000-0002-1495-7052}
\email{dlee105@ucsc.edu}
\affiliation{%
  \institution{University of California, Santa Cruz}
  \city{Santa Cruz}
  \state{California}
  \country{USA}
}

\author{Leilani H. Gilpin}
\orcid{0000-0002-9741-2014}
\email{lgilpin@ucsc.edu}
\affiliation{%
  \institution{University of California, Santa Cruz}
  \city{Santa Cruz}
  \state{California}
  \country{USA}
}

\renewcommand{\shortauthors}{Liu et al.}

\begin{abstract}
On a 3D-printed tactile map, a building felt under the finger is an anonymous shape: touch alone cannot tell which footprint is which, and a spoken description cannot reliably point to one shape at one place. We present a web-based tool that lets a sighted helper click to add on-shape Braille labels to an already-generated map model, downstream of the geometry generator so that whoever knows the reader and the local Braille standard does the labeling. The tool offers click-based OpenStreetMap matching, hand-editable abbreviation that shrinks a name to fit a footprint, and print-safe dot geometry with a review step that catches anomalies before printing. We demonstrate it on five printed maps of different place types, from a downtown core to a college campus and a small dining mall. In formative sessions in which ten BLV readers compared an unlabeled print with an annotated one, four read Braille fluently, so we treat Braille as one output among several rather than the only one. The tool's core is the link between coordinates, geometry, and a place's semantics, which can drive an audio readout or a non-Braille code. The tool is available at \url{https://leolee7.github.io/Annotate_Braille/}.
\end{abstract}



\maketitle

\begin{figure*}[t]
  \centering
  \includegraphics[width=0.8\linewidth, alt={A six-panel figure, all of the same college-campus map. (A): a photograph of a 3D-printed tactile map from Touch Mapper with raised building footprints and no labels. (B): a photograph of the same map after annotation, with raised Braille on the footprints. (C): a screenshot of the browser annotation editor working on that same map, with a grey 3D relief model on the left whose building rooftops are clickable, green footprints already annotated and one yellow footprint currently selected (cursor 1), a 2D OpenStreetMap reference map of the campus in the center where the annotator clicks the matching building (cursor 2), and a label panel on the right showing the English name Advanced Technology Center and the Braille code ATC. (D): a Braille legend table mapping codes such as AC, CC, ADM, ATC to building names. (E): a category-organized text description of the map with positions, coverage percentages, GPS, and OSM links. (F): a per-label JSON record grouped into geometry on the print model in millimeters, the Braille output, and OpenStreetMap geospatial semantics.}]{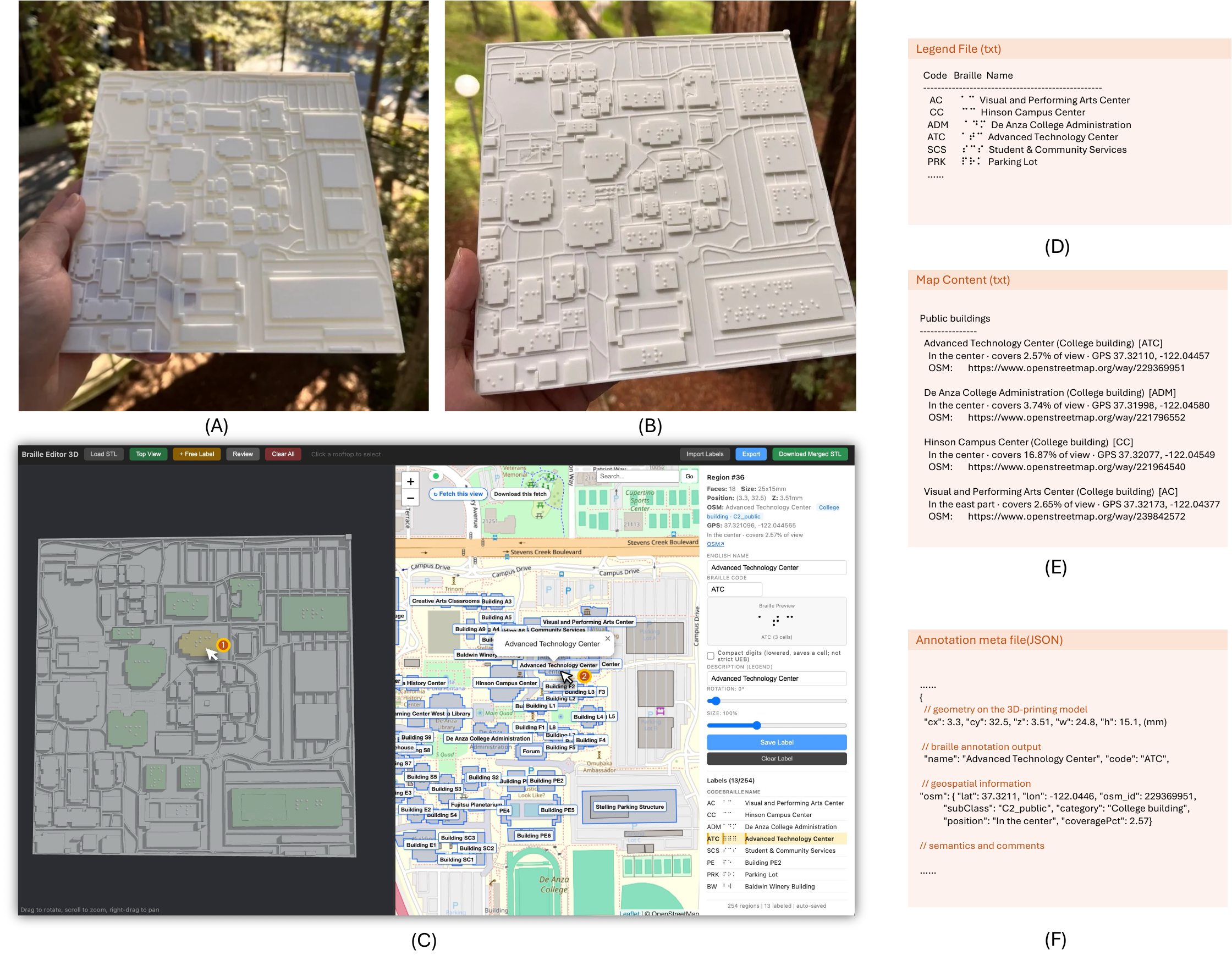}
  \caption{Our annotation tool applied to one college-campus map, from the unlabeled print to the exported files. (A) The Touch Mapper map printed without labels: anonymous relief shapes. (B) The same map after annotation: the footprints carry on-shape Braille. (C) The editor working on that same map (the STL behind A). Every building rooftop on the model is clickable: green marks footprints already annotated, and yellow marks the one currently selected (cursor~1). The annotator then picks the matching building on the retrieved OpenStreetMap reference (cursor~2) and labels it directly, and can reopen any label to edit it at any time. (D--F) Examples of the three files the tool exports for the map: a Braille legend (D), a category-organized map description (E), and a per-label JSON record that ties each shape's print-millimeter geometry to its OpenStreetMap semantics (F). Map data in (C) \copyright{} OpenStreetMap contributors, openstreetmap.org/copyright.}
  \Description{A six-panel figure, all of the same college-campus map. (A): a photograph of a 3D-printed tactile map from Touch Mapper with raised building footprints and no labels. (B): a photograph of the same map after annotation, with raised Braille on the footprints. (C): a screenshot of the browser annotation editor working on that same map, with a grey 3D relief model on the left whose building rooftops are clickable, green footprints already annotated and one yellow footprint currently selected (cursor 1), a 2D OpenStreetMap reference map of the campus in the center where the annotator clicks the matching building (cursor 2), and a label panel on the right showing the English name Advanced Technology Center and the Braille code ATC. (D): a Braille legend table mapping codes such as AC, CC, ADM, ATC to building names. (E): a category-organized text description of the map with positions, coverage percentages, GPS, and OSM links. (F): a per-label JSON record grouped into geometry on the print model in millimeters, the Braille output, and OpenStreetMap geospatial semantics.}
  \label{fig:teaser}
\end{figure*}

\section{Introduction}

Blind and Low-Vision (BLV) people use tactile maps to build spatial knowledge of neighborhoods and buildings~\cite{brock2015interactive,taylor2016customizable}, and low-cost printing has made such maps inexpensive to produce from open data~\cite{touchmapper}. But without labels, a tactile map is just a field of anonymous shapes. When a reader runs a finger across a printed block, they feel footprints, streets, and open space, yet nothing on the surface says which footprint is the library and which is the station. As one user of Touch Mapper reported on its issue tracker, a reader ``had to ask for the name of the street that she was touching to help her figure out the map''~\cite{touchmapper_issue7}. The missing layer is on-shape symbolic identity: a readable Braille label on the shape itself gives the reader a finger-locatable anchor they can read by touch.

Existing generation pipelines rarely carry this label layer. Automatic label placement has been built for flat map sheets: a script pipeline for swell-paper city maps places Braille and print labels, and abbreviates street names too long to fit in Braille, and its authors report that complex cases still need human inspection and editing~\cite{stampach2016automated}. Reviews add that production as a whole is not automated end to end and still depends on a team of specialists~\cite{wabinski2019automatic,mukhiddinov2021systematic}. Recent automation targets the map geometry rather than the labels: a proof-of-concept computer-vision model translates rendered map images into tactile line and area features, explicitly leaving text and Braille to a separate component~\cite{hobson2024gan}. Commercial workflows take a different route: Esri's tutorial places Braille with a font and non-raising print, with the legend on a separate page, for swell paper rather than 3D prints~\cite{esri_tactilemap_tutorial}. What is missing is a way to add or revise those labels after the model exists, for a particular reader.

The labeling that automation skips is a human-judgment task. It depends on knowing the reader and the local Braille standard, choosing a readable name or abbreviation, and matching it to the right shape, which is why practitioners argue that labeling should stay with a person rather than be built into the generator (Section~\ref{sec:system}). We therefore support this step instead of trying to remove it. This demo presents a web-based, click-to-annotate tool that lets a sighted helper place Braille labels on an already-generated 3D-print model, decoupled from the geometry generator so that whoever knows the reader and the local standard does the labeling~\cite{touchmapper_issue7}, an arrangement consistent with evidence that tactile maps must be tailored to individual users' needs~\cite{vanaltena2023tailoring}. The tool records each label as a structured entry that ties a place's name and OpenStreetMap identity to its position on the print, so the labeling effort can later feed outputs beyond Braille (Fig.~\ref{fig:bridge}).

Our contributions are (1) a print-validated, semi-automatic tool that lets a sighted helper tailor the labels to a particular reader, closing the on-shape labeling gap downstream of geometry, with click-based OpenStreetMap matching and hand-editable abbreviation that shrinks a name to fit a footprint; (2) print-safe Braille dot geometry with a review step that flags anomalies before printing; and (3) an editable per-label record that separates print-specific geometry from reusable semantics, so the same annotation can also drive outputs for readers who do not read Braille. Voice input and an annotation assistant remain future work.

\section{Related Work}
\label{sec:related}

\textbf{Tactile map generation and the labeling bottleneck.}
Tactile cartography has conventions for touch-based symbol design and for generalizing detail to a readable density~\cite{vanaltena2025tactile}, and the label layer needs its own rule as scale shrinks, which our name-to-code fallback supplies. Studies of blind readers' needs stress a clear title, an overview, a legend, and distinguishable icons on a sheet with limited space~\cite{vanaltena2023tailoring}. Generation tools mostly separate geometry from text: Touch Mapper ships printable geometry but no Braille~\cite{touchmapper}, and a recent proof-of-concept computer-vision model automates the tactile feature layer, as a 2D image, but not the labels~\cite{hobson2024gan}. One rule-based pipeline does render Braille labels into a 3D-printable model automatically~\cite{gotzelmann2014towards}, which is the arrangement the practitioners in the issue thread argue against: the labels are fixed by the generator's rule set, so they cannot be matched to the Braille standard a particular reader uses or to the features that reader wants named. Reviews report that tactile map production is not automated end to end and still depends on a team of specialists~\cite{wabinski2019automatic,mukhiddinov2021systematic}. Automatic label placement, including the abbreviation of names too long for Braille, has been developed for flat swell-paper map sheets, where complex cases are still passed back to a person~\cite{stampach2016automated}. Tactile Vega-Lite automates Braille for charts, which have a regular grid that map footprints lack~\cite{chen2025tactilevegalite}, and Esri's tutorial keeps Braille a font-and-layout step on swell paper~\cite{esri_tactilemap_tutorial}. We instead annotate already-generated 3D geometry with print-safe on-shape Braille and export a reusable per-label record.

\textbf{Device- and agent-based reading aids.}
A second line of work makes an inert artifact readable or queryable by adding a layer that works only while its hardware or input image is present: audio registered to touch through instrumented overlays, touchscreens, or capacitive and camera sensing~\cite{brock2015interactive,gotzelmann2016lucentmaps,taylor2016customizable,gotzelmann2016capmaps,shi2017markit}, and conversational agents over tactile models~\cite{cavazos2019jido,reinders2020hey}, touch-explored digital maps~\cite{liu2026touchingspace}, or digital charts~\cite{gorniak2024vizability}, alongside benchmarks of the visual questions blind people ask~\cite{gurari2018vizwiz,chen2022grounding} and a proposed agent architecture over street-level imagery~\cite{froehlich2025geospatial}. These systems answer richer queries than a printed label can, but the identity of a shape lives in the companion device rather than on the shape. We instead make the printed shape self-describing through on-shape Braille, read by touch without the powered hardware those systems need. The per-label record our tool exports is the grounding such an agent would require.

\textbf{Braille on a printed surface.}
A third literature specifies the label itself rather than where it goes. Evidence on tactile symbol design shows that a symbol's height, not spacing alone, governs whether it reads, and that differentiating height lets the minimum horizontal spacing between symbols drop~\cite{wabinski2022height}, while standards specify permitted ranges for cell and dot geometry on physical objects~\cite{iso17049}. This work tells an annotator what a legible symbol must be, and leaves open how to place one on a particular model without breaking it, which our dot geometry and review step handle.

\section{System}
\label{sec:system}

\begin{figure}[t]
    \centering
    \includegraphics[width=1\linewidth, alt={A diagram with a central box labeled Our platform, joined by double-headed arrows to four boxes: Real World, geospatial information; 3D-printed artifact, tangible annotations; Sighted Annotators, easy to label and audit; and BLV user, customizable Braille and symbol support. Below them a wide bar labeled Shareable Annotations, reuse to other applications, holds four items: an online worldwide Braille-OSM library, support for LLMs and agents, an audio readout, and an ellipsis for further uses.}]{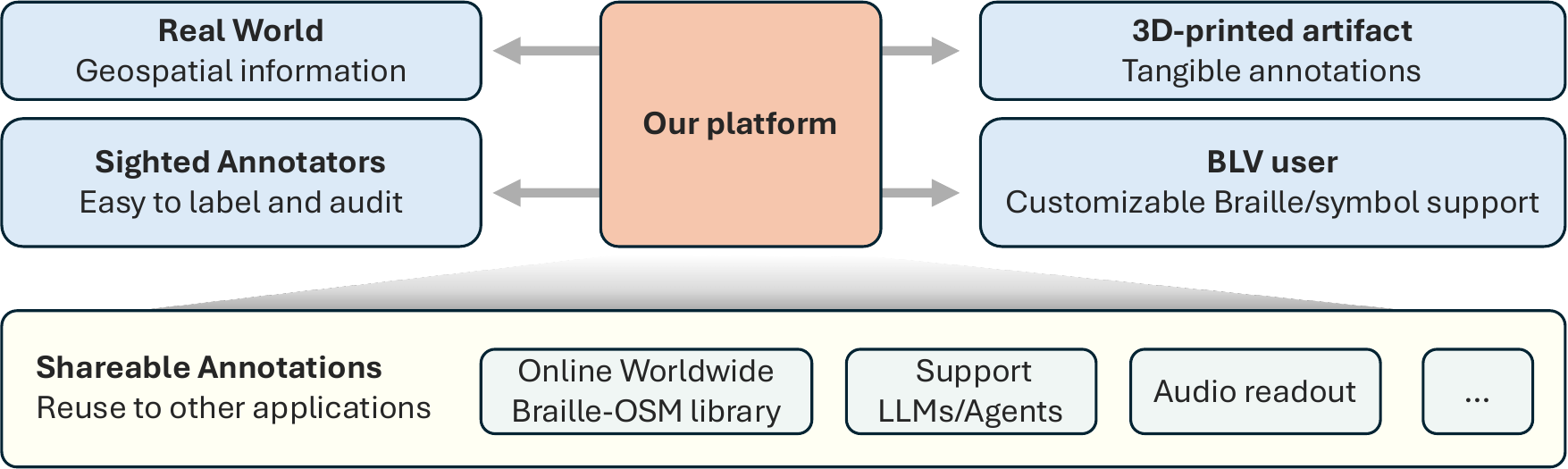}
    \caption{The tool, drawn as a bridge between four parts: the real world (geospatial information), the 3D-printed artifact, the sighted annotator, and the BLV reader. The per-label annotations it exports are reusable beyond one print, across a Braille-OSM library, LLMs and agents, and an audio readout.}
    \Description{A diagram with a central box labeled Our platform, joined by double-headed arrows to four boxes: Real World, geospatial information; 3D-printed artifact, tangible annotations; Sighted Annotators, easy to label and audit; and BLV user, customizable Braille and symbol support. Below them a wide bar labeled Shareable Annotations, reuse to other applications, holds four items: an online worldwide Braille-OSM library, support for LLMs and agents, an audio readout, and an ellipsis for further uses.}
    \label{fig:bridge}
\end{figure}

The tool brings four parts together in one view: OpenStreetMap data (coordinates, names, and categories), the 3D-printed STL model in millimeters, the sighted annotator, and the BLV reader (Fig.~\ref{fig:bridge}). The annotator places a label with a click instead of measuring coordinates. The tool runs in a browser and exports a print-ready annotated STL and a reusable per-label record.

We keep labeling as a downstream stage rather than pushing Braille into the geometry tool that produces the model~\cite{touchmapper}. The decision comes from a public issue thread on Touch Mapper's tracker, open since 2018, where its users and its maintainer work out how labels should be added~\cite{touchmapper_issue7}. An assistive-technology professional there argued that baking Braille in at the source is impractical, since many Braille standards coexist even within English and a maintainer cannot know which points of interest each reader wants labeled, and proposed instead that the tool export an editable model so that those helping a specific reader can add the Braille that reader needs. We built our tool along that line and posted it to the thread, and the maintainer judged manual placement better than automatic. Building the tool we reached the same position: manual placement was more reliable for matching shapes to entities, choosing abbreviations, and handling edge cases such as footprints too small to label, open areas, and overlapping geometry. The one pipeline that does abbreviate and place Braille labels automatically still hands complex cases back to a person~\cite{stampach2016automated}. We therefore keep a human in the loop and reduce the effort it takes.

\subsection{Click-to-annotate workflow}
\label{subsec:annotate}

The editor shows the 3D model beside a 2D reference map (Fig.~\ref{fig:teaser}C) and makes both clickable: it detects each rooftop on the model as a selectable region, and one click on ``Fetch this view'' pulls every OpenStreetMap building footprint in the current map view on demand, each a clickable polygon. The annotator pairs a shape with its map entity in either order, clicking the rooftop and the matching footprint (the two cursors in Fig.~\ref{fig:teaser}C). The tool then fills in the name, a short description, and the Braille code, with per-feature metadata in Touch Mapper's scheme: a functional classification, GPS, a spatial hint, and OSM and Wikidata links. When a building has no name, a point of interest inside its footprint supplies one. Confirmation stays manual because shape-to-entity alignment is not reliable enough to accept unchecked.

Since a name rarely fits a footprint, the editor proposes a hand-editable abbreviation from three rule tiers: function keywords (\emph{Advanced Technology Center} $\rightarrow$ ATC), campus codes (\emph{Building S5} $\rightarrow$ S5), and filtered initials (\emph{Neiman Marcus} $\rightarrow$ NM), with the expansion recorded in an exported legend key. By default the codes follow UEB conventions, including the number indicator that distinguishes digits from a to j. A compact mode is available for tight footprints.

\subsection{Print-safe output and reusable record}
\label{subsec:output}

To survive 3D printing, we optimized the Braille dots. Our early full-sphere dots pierced thin shells, so each dot is now a hemisphere on a flat disc, with an adaptive base slab that keeps at least 1\,mm of material under it. Cell and dot dimensions, including dot height, stay within the ranges ISO 17049 specifies~\cite{iso17049}. A Review mode flags print problems before export, such as dots that pierce the base and labels that overlap or float.

The tool is an open-source page that runs entirely in the browser, so a helper needs nothing installed and no account. It exports the merged STL together with a per-label record that keeps print-specific geometry (position, rotation, scale) separate from reusable semantics (the name, its Braille, classification, and OSM links). Because the two are separate, a record can be revised or recomposed across sheets without redoing the annotation.

\section{Workflow and Demonstration}
\label{sec:demo}

\textbf{From a request to a readable map.} The work starts with a place a BLV user wants to know. A sighted helper generates that area in Touch Mapper~\cite{touchmapper},\footnote{\url{https://touch-mapper.org/}} downloads the STL, and loads it into our editor (Fig.~\ref{fig:teaser}C). Clicking a footprint on the 2D map and on the 3D model pairs them and fills in the name and Braille code. The helper adjusts placement, runs Review to catch a dot that would pierce the base or collide with a neighbor, and exports. Because the map and the model stay linked, the helper does not measure coordinates or compare against a separate static map. The user comes away with a printed map whose shapes are no longer anonymous, so that a Braille reader can identify and locate buildings by touch instead of asking a companion for each one.

\textbf{Reuse beyond one print.} The exported per-label record ties each label to its OpenStreetMap entity, its position on the print, and its Braille, keeping geometry and meaning separate. The same record can drive an exported legend today or a neutral symbol with a press-to-hear readout later, so the labeling carries across prints and across output forms.

\textbf{How we gathered feedback.} Our design draws on two inputs. The first is the issue-thread discussion with Touch Mapper's users and maintainer described in Section~\ref{sec:system}. The second is hands-on feedback from readers. We annotated five printed maps of different place types, among them a downtown core, a college campus, and a small dining mall, so the workflow is not specific to a single kind of site (Fig.~\ref{fig:teaser}A and~B show one campus map before and after annotation). We annotated the maps, and three research assistants who had not been involved in building the tool also used it. Ten BLV readers, one of them a Braille instructor, then handled the prints at a blind services center, as one segment of a longer session on spatial access. Each reader was given the unlabeled Touch Mapper print and the annotated print of the same area side by side and was asked to locate the Braille, read the codes, and say what each one stood for, followed by open questions on dot feel, spacing, and what the labels would need in order to be usable. Sessions were audio-recorded and transcribed, and we reviewed the transcripts and grouped the recurring points into the three considerations below. The study was approved by our institution's review board, and participants gave informed consent. The feedback is formative.

\textbf{Three considerations for on-shape labeling.} First, on-shape labels change what the map affords. One reader said unlabeled tactile maps make sense only with help: ``most of the time they don't make sense until someone explains it to you.'' The annotated map is meant to reduce that dependence, since Braille readers read most labels without extra training. This turns anonymous shapes into named ones, a shift from \emph{where} to \emph{what}. Second, the codes are legible but not self-explaining. Readers read the letters easily, calling the dots ``pretty good'' and saying they ``stand out nicely,'' yet could not always say what a code stood for. The gap is the abbreviation, and the instructor asked for a key, which our exported legend provides. Third, Braille cannot be assumed as the sole output: four of the ten read it fluently, and the rest pointed toward a press-to-hear or non-Braille form, which the same per-label record can drive. The Braille instructor herself saw Braille giving way to digital formats for reading, while remaining essential for everyday labels such as clothing and pill bottles. A map's labels should therefore not be tied to Braille alone.

\section{Discussion and Future Work}
\label{sec:discussion}

\textbf{Limitations.} The tool still needs a sighted annotator to confirm matches and abbreviations. Much of the Braille layer also has room to improve: the abbreviation scheme, support for contracted Braille, and per-reader cell sizing could all be stronger, and the legend could do more to resolve ambiguous codes. One reader who read the codes without difficulty still asked for them ``a little bigger.'' The labeling is tied to printed Braille for now, whereas readers pointed toward symbol and audio forms that decouple identity from print, which we have only begun to explore. Our evaluation so far is formative. A natural next step is to work with the teachers and assistive-technology professionals who would do the annotating.

\textbf{Future work.} Beyond Braille, readers pointed to a press-to-hear audio readout or a compact non-Braille symbol code, both driven by the same per-label record. A volunteer had walked one reader to the nearest bus stop on a first visit. That reader said the annotated print ``would have been very helpful for me,'' and pictured one kept at a services center for newcomers. Because every record is keyed to an OpenStreetMap entity, the annotations could also accumulate into a shared layer of Braille and place semantics over OpenStreetMap, so that whoever prints the same blocks next starts from what an earlier helper already named. The same records could ground a conversational agent in the entities under a reader's finger rather than in a photo~\cite{gurari2018vizwiz,chen2022grounding,liu2024rightthisway,froehlich2025geospatial}, extending the audio-haptic map exploration we have built on a trackpad~\cite{liu2026touchingspace}. We plan a controlled comparison of whether on-shape labels improve independent locating, against shapes-only and audio-only reading.

\textbf{Conclusion.} We presented a semi-automatic tool that prepares Braille for 3D-printed tactile maps, keeping labeling downstream of geometry, and a per-label record that links coordinates, geometry, and place semantics, which sets up the audio, symbol, and agent outputs we pursue next.

\section{GenAI Usage Disclosure}
At run time, the tool does not rely on generative AI: its labels come from real OpenStreetMap data, and the Braille codes, abbreviations, and print-safe geometry are produced by deterministic rules rather than GenAI models. The authors reviewed all AI-assisted output and take full responsibility for the content, including the ideas, design, and results.

\begin{acks}
This material is based upon work supported by the Air Force Office of Scientific Research under
award number FA9550-24-1-0149. 

We thank Prof. Roberto Manduchi for his feedback and guidance,
Gauri Jain, Sanya Bhatia, and Rhea Susarla for their help with the tool, and the Touch Mapper maintainer and
issue-thread contributors whose discussion shaped the design. The maps were printed with
support from Baskin Engineering Lab Support, the Slugworks makerspace, and the University
Library's Digital Scholarship Innovation Studio at UC Santa Cruz. We thank the Vista Center for
the Blind and Visually Impaired, and its Santa Cruz, San Jose, and Palo Alto offices, for the
use of their space and for connecting us with participants. We especially thank the
participants in our pilot and formative sessions, and the blind and low-vision community more
broadly, for the time and the insight that shaped this work.
\end{acks}

\bibliographystyle{ACM-Reference-Format}
\bibliography{reference}

\end{document}